\documentclass[aip,preprint]{revtex4-1}
\usepackage{graphicx}
\usepackage{amsfonts}
\usepackage{amsmath}
\usepackage{color}
\usepackage{color} 
\usepackage{multirow}
\makeatletter

\newcommand{\Rmnum}[1]{\expandafter\@slowromancap\romannumeral #1@}
\makeatother

\begin{document}

%\preprint{AIP/123-QED}

\title{Observation of spin glass state in weakly ferromagnetic Sr$_2$FeCoO$_6$ double perovskite}
\author{Pradheesh R.}
\affiliation{Low Temperature Physics Laboratory, Department of Physics, Indian Institute of Technology Madras, Chennai 600036, India.}
\author{Harikrishnan S. Nair}
\affiliation{J\"{u}lich Center for Neutron Sciences-2/Peter Gr$\ddot{u}$nberg Institute-4, Forschungszentrum J\"{u}lich, 52425 J\"{u}lich, Germany}
\author{C. M. N. Kumar}
\affiliation{J\"{u}lich Center for Neutron Sciences-2/Peter Gr$\ddot{u}$nberg Institute-4, Forschungszentrum J\"{u}lich, 52425 J\"{u}lich, Germany}
\author{Jagat Lamsal}
\affiliation{Department of Physics and Astronomy, University of Missouri-Columbia, Missouri 65211, USA.}
\author{R. Nirmala}
\affiliation{Low Temperature Physics Laboratory, Department of Physics, Indian Institute of Technology Madras, Chennai 600036, India.}
\author{P. N. Santhosh}
\affiliation{Low Temperature Physics Laboratory, Department of Physics, Indian Institute of Technology Madras, Chennai 600036, India.}
\author{W. B. Yelon}
\affiliation{Materials Research Center and Department of Chemistry, Missouri University of Science and Technology, Rolla, Missouri 65409, USA.}
\author{S. K. Malik}
\affiliation{Departmento de f$\acute{i}$sica Te$\acute{o}$rica e Experimental, Natal-RN, 59072-970, Brazil.}
\author{V. Sankaranarayanan$^*$}
\affiliation{Low Temperature Physics Laboratory, Department of Physics, Indian Institute of Technology Madras, Chennai 600036, India.}
\author{K. Sethupathi$^*$}
\affiliation{Low Temperature Physics Laboratory, Department of Physics, Indian Institute of Technology Madras, Chennai 600036, India.}
%\email{vsn@physics.iitm.ac.in, ksethu@physics.iitm.ac.in}
\date{\today}
\begin{abstract}
We report the observation of spin glass state
%with a reduced transition
in the double perovskite oxide Sr$_{2}$FeCoO$_{6}$
prepared through sol-gel technique.
%
%with a transition temperature, $T_f \sim$ 75~K, through
%magnetic measurements.
%
Initial structural studies using x rays reveal that the compound
crystallizes in tetragonal $I 4/m$ structure with lattice parameters,
$a$ = 5.4609(2)~\mbox{\AA} and $c$ = 7.7113(7)~\mbox{\AA}.
The temperature dependent powder x ray studies reveal
no structural phase transition in the temperature range 10 -- 300~K.
However, the unit cell volume shows an anomaly coinciding with the
magnetic transition temperature thereby suggesting a close connection between
lattice and magnetism.
Neutron diffraction studies and subsequent bond valence sums
analysis show that in Sr$_{2}$FeCoO$_{6}$,
the $B$ site is randomly occupied by Fe and Co
in the mixed valence states of
Fe$^{3+}$/Fe$^{4+}$ and Co$^{3+}$/Co$^{4+}$.
The random occupancy and mixed valence
sets the stage for inhomogeneous magnetic exchange interactions and
in turn, for the spin glass like state in this double
perovskite which is observed as an irreversibility
in temperature dependent dc magnetization at $T_f\sim$ 75~K.
Thermal hysteresis observed in the magnetization
profile of Sr$_{2}$FeCoO$_{6}$ is indicative of the mixed
magnetic phases present.
The dynamic magnetic susceptibility displays characteristic
frequency dependence and confirms the spin glass nature of this material.
Dynamical scaling analysis of $\chi'(T)$ yields a critical temperature $T_{ct}$ = 75.14(8)~K
and an exponent $z\nu$ = 6.2(2) typical for spin glasses.
The signature of presence of mixed magnetic interactions
is obtained from the thermal hysteresis in magnetization of Sr$_{2}$FeCoO$_{6}$.
Combining the neutron and magnetization results of Sr$_2$FeCoO$_6$, we deduce
the spin states of Fe to be in low spin while that of Co to be in low spin 
and intermediate spin.
\end{abstract}
%
%=======================================================%
\pacs{75.50.Lk, 75.47.Lx, 75.50.-y}
\maketitle
%
%=======================================================%
%
\section{Introduction}
Ferromagnetic double perovskite oxides of general formula $A_{2}BB'O_{6}$
($A$ = divalent alkaline earth;  $B$ and $B'$ = transition metals)
have been extensively studied as candidates for magnetoresistive materials.
\cite{serratejpcm_19_023201_2007,sarmassc_114_465_2000}
The crystallographic ordering of the $B$ cations in the double perovskites plays an important role in
realizing novel magnetic and transport properties including magnetoresistance.
\cite{sarmassc_114_465_2000}
Depending on the crystallographic arrangement of $B$ cations,
double perovskites are classified as random, rocksalt or layered.
\cite{anderson_1993}
For example, double perovskites like Ca$_2$FeMoO$_6$  have a rocksalt arrangement
\cite{gopalakrishnan_2000}
while  PrBaCo$_2$O$_{5+\delta}$ adopts a layered structure.
\cite{streule_2005}
Though $B$ site ordered double perovskites are of interest for magnetoresistive properties
and ferromagnetism, it has been observed that comparable ionic radii and oxidation states
of different cations at the $B$ site can lead to mixed crystallographic occupation and result
in what is known as antisite disorder.
\cite{anderson_1993}
Depending on the type, size and charge of the cations present at the
$B$ site, different magnetic properties varying from antiferromagnetism (AFM)
\cite{cussenjmc_7_459_1997,kobayashijmmm_218_17_2000}
to spin glass (SG)
\cite{battlejssc_78_281_1989,poddarjap_106_073908_2009evidence}
have been reported for double perovskites.
Most of the Fe based double perovskites display a high magnetic transition
temperature, which is surprising given the fact that the 
Fe ions are separated far apart in this compound.
\cite{sarmassc_114_465_2000}
However, a few exceptions to this, like Sr$_2$FeWO$_6$, have been reported,
with a low transition temperatures of $\approx$ 37~K.
\cite{kawanakaphysica_281_518_2000iron}
The correlation between magnetoresistance and magnetism in these
compounds and the site disorder is clear from the study comparing the
properties of ordered and disordered Sr$_2$FeMoO$_6$.
\cite{navarrojpcm_13_8481_2001}
The site disorder can partially or completely
destroy the $B$ site ordering of cations
and can influence the physical properties like magnitude of magnetization,
Curie temperature and low-field magnetoresistance.
\cite{balcellsapl_78_781_2001cationic,garciahernandezprl_86_2443_2001}
It has been found that the site disorder can be influenced by preparative
conditions such as heat treatment temperature
\cite{balcellsapl_78_781_2001cationic,sakumajap_93_2816_2003},
treatment time
\cite{sakumajap_93_2816_2003} etc.,
thereby making it possible to tune the magnetic phases
of $A_{2}BB'O_{6}$ through a suitable selection of $B/B'$ cations
and preparative conditions.
Though double perovskite oxides in the $A_2BB'$O$_6$ family have been actively investigated,
there are only a few reports on the Co based double perovskite, Sr$_2$FeCoO$_6$.
\cite{maignan_3_57_2001,bezdicka_91_501_1994}
Earlier studies on the chemically similar single perovskite SrFe$_{1-x}$Co$_{x}$O$_{3}$
reported the $x$ = 0.5 composition to be ferromagnetic with a high
$T_{c}$ of 340~K and saturation magnetic moment of 3~$\mu_{B}$.
\cite{takeda_33_973_1972,kawasakijssc_121_174_1996}
Meanwhile, Maignan $et~al.,$ reported SrFe$_{0.5}$Co$_{0.5}$O$_{3}$
to be ferromagnetic below $T_{c}\sim$ 200~K,
with a saturation magnetization of 1.5~$\mu_{B}$ at 5~K, in 1.45~T.
\cite{maignan_3_57_2001}
{\it Ab~initio} band structure calculations on the double perovskite Sr$_2$FeCoO$_{6}$
showed that Co and Fe make comparable contributions to ferromagnetism.
\cite{bannikov2008}
The Co$^{4+}$($d^5\bar{L}$) and Fe$^{4+}$($d^4\bar{L}$) in high spin (HS) states
can lead to metallicity and ferromagnetism in these systems.
However, owing to the comparable ionic radii and
valence states of Fe and Co (both in 4+ state), it is surprising
that Co doped SrFeO$_3$ showed ferromagnetism.
The absence of a linear $B$--O--$B'$--O--$B$ chain consequent to tilts in the $B$O$_6$
and $B'$O$_6$ octahedra can cause 90$^{\circ}$ and 180$^{\circ}$ superexchange interactions.
\cite{poddar_106_2009,Elad_2003}
Moreover, the comparable ionic radii of the $B$ site cations
combined with the antisite disorder can lead to magnetic frustration in the double perovskites.
Cluster glass phenomenon in Sr$_2$Mn$_{1-x}$Fe$_x$MoO$_6$
\cite{poddar_106_2009}
was attributed to the local magnetic frustration
developed due to the competing nearest neighbour ($NN$)
and next nearest neighbour ($NNN$) superexchange interactions.
Incompatible superexchange interactions and magnetic frustration due to the
site disorder of the $B$ and $B'$ cations were observed also in Sr$_2$FeTiO$_{6-\delta}$.
\cite{gibb_2_1992}
These results point to the fact that mixed valence state
of the $B$ and $B'$ ions and their crystallographic disorder
can lead to a spin glass state as it would create {\it mixed interactions}
and {\it randomness}.
\cite{mydosh_1993}
In the present study on Sr$_{2}$FeCoO$_{6}$,
through detailed structural and magnetic
investigations, we observe an inhomogeneous magnetic
state with the characteristics of a spin glass phase
arising from incompatible magnetic interactions and disorder.
\section{Experimental Details}
Polycrystalline samples of Sr$_2$FeCoO$_6$
were prepared following sol-gel method as described elsewhere
\cite{deacprb65_174426_2002} 
but, drying the gel at 120$^{\circ}$C and
performing the final sintering at 1050$^{\circ}$C for 36~h.
X-ray powder diffraction (XRD) patterns were recorded in the
temperature range of 10 - 300~K using 
a Huber diffractometer in Guinier geometry (Mo K$_{\alpha}$).
Neutron powder diffraction (NPD) measurement at 300~K was performed at the University
of Missouri Research Reactor (MURR) with neutron wavelength of 1.4789 \mbox{\AA} employing position sensitive
detector.
The crystal structure was refined by Rietveld method
\cite{rietveld}
using FULLPROF program. \cite{carvajal}
DC and AC magnetic measurements were carried out in a
commercial SQUID magnetometer and physical property measurement
system (both M/s Quantum Design, USA).
DC magnetization in field cooled (FC) and zero-field cooled (ZFC) cycles were
performed at different applied fields from 100 to 50~kOe.
AC susceptibility was measured using SQUID magnetometer in an ac field of 3 Oe with frequencies of
$33$, $133$, $337$, $667$, $967$, $1333$ Hz.
\section{Results and discussion}
\begin{figure}%[!h]
\centering
\includegraphics[scale=0.40]{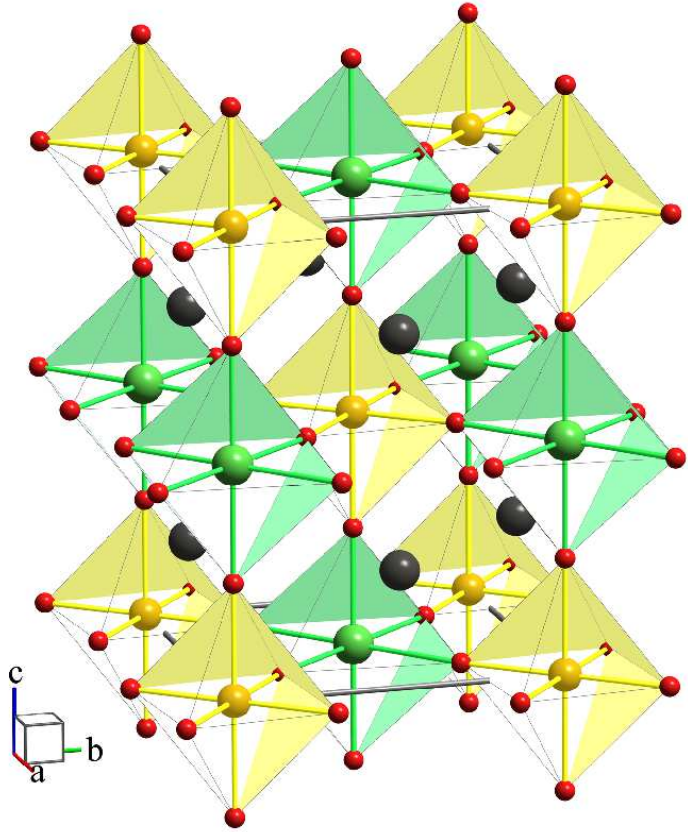}
\caption{The tetragonal structure of the double perovskite Sr$_2$FeCoO$_6$. The oxygen atoms
are represented by red spheres while Co are green, Fe are yellow and Sr are dark grey. The CoO$_6$ and FeO$_6$ octahedra are also indicated.
For succintness the structure shown is of the $B$ site ordered type. }
\label{structure}
\end{figure}

\begin{figure}%[!h]
\centering
\includegraphics[scale=0.55]{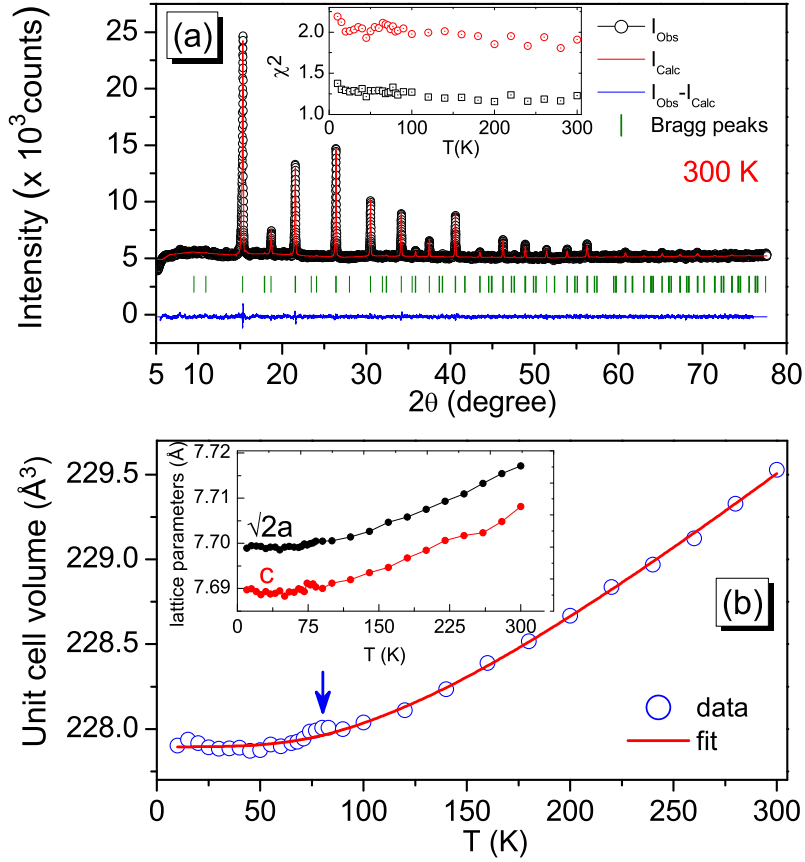}
\caption{(a) X-ray diffraction pattern of Sr$_{2}$FeCoO$_{6}$ at 300~K
refined in tetragonal $I 4/m$ space group. 
Quality measures of refinement are R$_{wP}$ = 1.52 $\%$, R$_{P}$ = 1.18 $\%$, $\chi^{2}$ = 1.23.
(b) Shows the analysis of unit cell volume using Gr\"{u}neisen approximation.
Note the anomaly in volume indicated by arrow. The inset shows temperature
evolution of lattice parameters $\sqrt{2}a$ and $c$ 
(the error bars were smaller than the size of the data markers).}
\label{xrd}
\end{figure}

\begin{figure}[!h]
\centering
\includegraphics[scale=0.55]{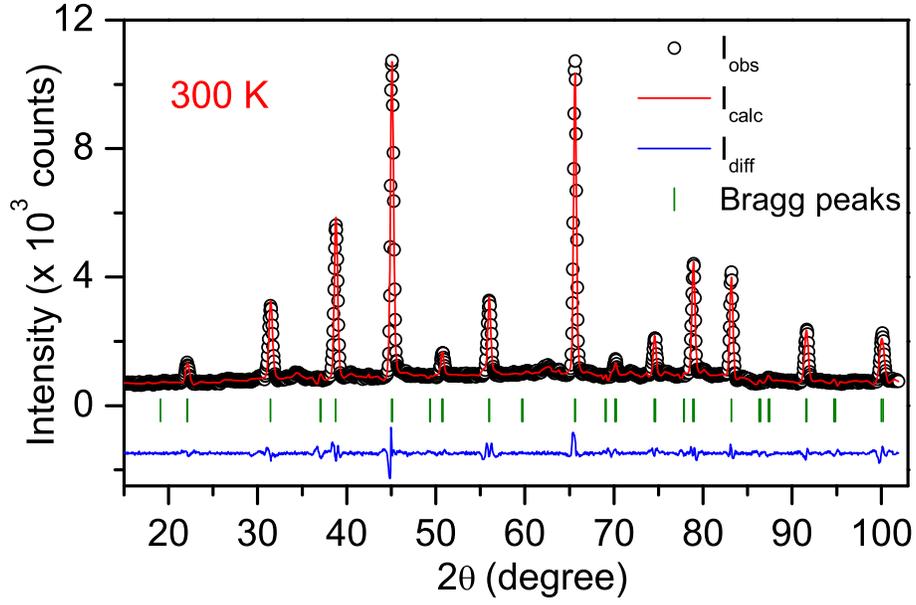}
\caption{(a) Observed (I$_{obs}$), calculated (I$_{calc}$) and difference (I$_{diff}$) profiles of
neutron diffraction pattern of Sr$_{2}$FeCoO$_{6}$ at 300~K refined in tetragonal $I 4/m$ space group. 
Vertical markers correspond to Bragg peaks.
Quality measures of refinement are R$_{wP}$ = 17.6 $\%$, R$_{P}$ = 14.6 $\%$, $\chi^{2}$ = 3.38.}
\label{ND}
\end{figure}

\begin{figure}[!ht]
\centering
\includegraphics[scale=0.55]{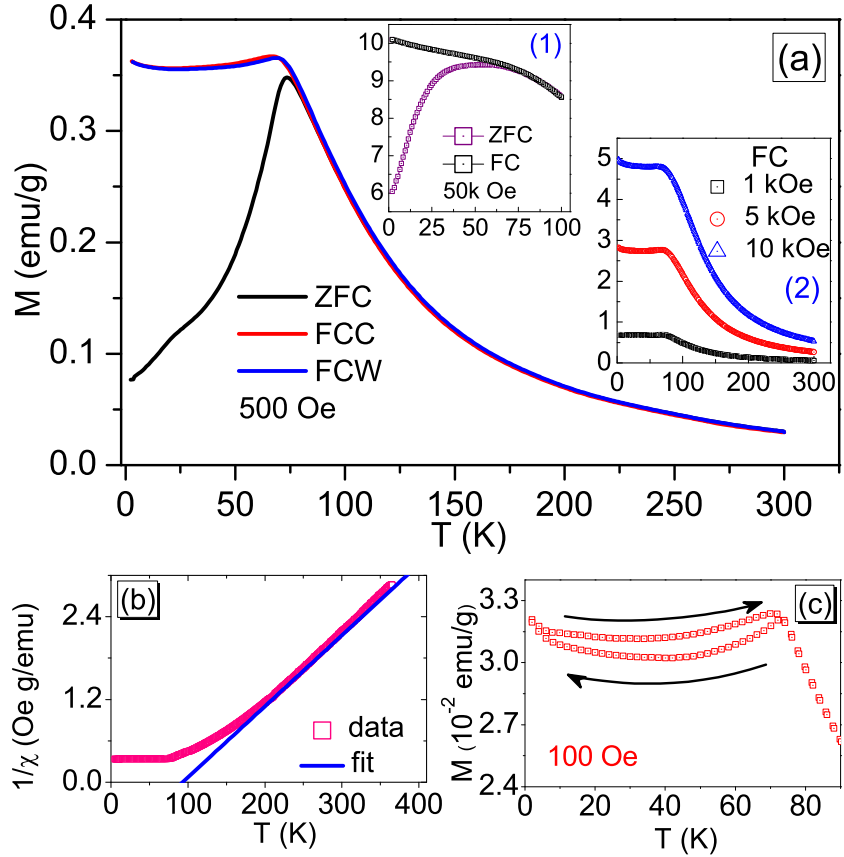}
\caption{(a) ZFC and FC (cooling\slash warming) cycles of magnetization of Sr$_2$FeCoO$_6$
as a function of temperature at 500 Oe.
Inset (1) shows the ZFC/FC curves at 50 kOe and (2) FC curves at different applied fields. 
(b) Temperature dependence of inverse magnetic susceptibility  along with Curie-Weiss fit.
(c) Thermal hysteresis observed in FC arm at 100~Oe.}
\label{MT}
\end{figure}

\begin{figure}[!h]
\centering
\includegraphics[scale=0.55]{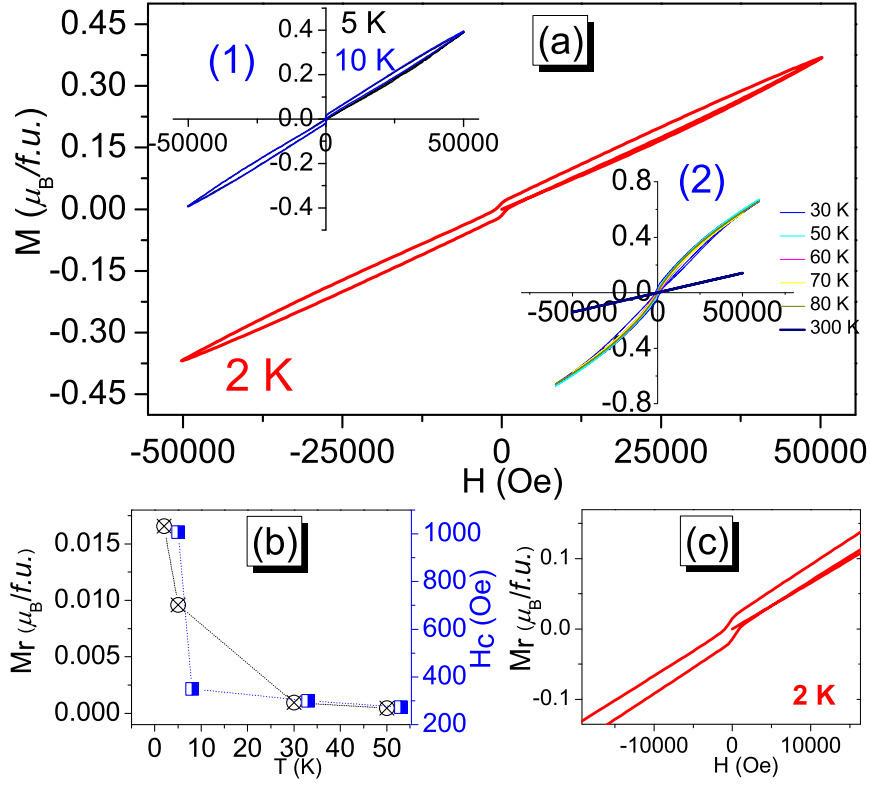}
\caption{(a) Isothermal magnetization of Sr$_2$FeCoO$_6$ at 2~K. Insets
(1) and (2) present isotherms at elevated temperatures. 
(b) Variation of remnant magnetization, $M_r$ and coercive field, $H_c$ with temperature.
(c) An enlarged view of the magnetization isotherms at 2~K for low fields.}
\label{MH}
\end{figure}

\begin{figure}[!h]
\centering
\includegraphics[scale=0.65]{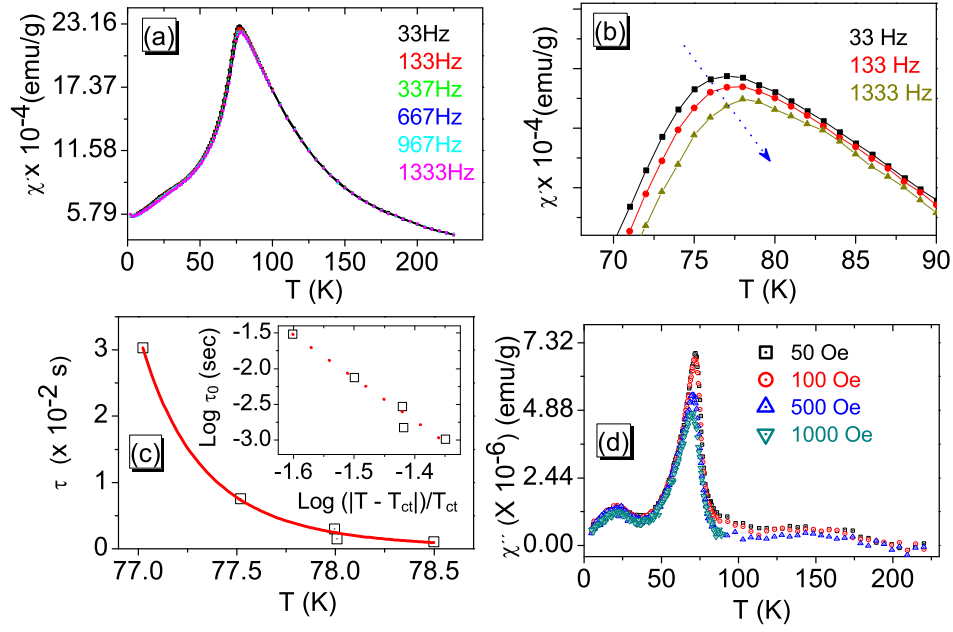}
\caption{(a) Temperature dependence of AC susceptibility 
of Sr$_2$FeCoO$_6$ at different frequencies.
(b) The peak in $\chi'(T)\sim$75~K which shifts to higher 
temperature as the frequency is increased, signifying glassy magnetism.
(c) The fit using the power law for critical slowing down, where the best fit 
gave $z\nu = 6.2$ , $\tau_{0} = 10^{-12}$ s. The inset presents the
same plot but in log-log scale.
(d) Displays $\chi''(T)$ at different values of superimposed magnetic fields.}
\label{acchi}
\end{figure}

%\subsection{Crystal structure}
%
From the powder XRD data, the crystal structure of Sr$_2$FeCoO$_6$ was
refined in the tetragonal space group $I 4/m$ (No. 87)
with lattice constants
$a$ = 5.4568(2)\mbox{\AA} and $c$ = 7.7082(4)~\mbox{\AA}.
A pictorial representation of the Sr$_2$FeCoO$_6$ double perovskite
unit cell is shown in Fig~\ref{structure}.
Even though, following the earlier reports on SrFe$_{1-x}$Co$_x$O$_3$ \cite{maignan_3_57_2001}
a refinement in cubic $Pm3m$ space group was undertaken initially,
a faithful fit to XRD data was achieved using the tetragonal space group.
The results of our Rietveld analysis are presented in Fig~\ref{xrd} (a)
where the observed xrd pattern at 300~K, the calculated profile, difference profile
and the Bragg peaks are shown.
The inset of Fig~\ref{xrd} (a) compares the $\chi^2$ (goodness-of-fit) of the refinement
from $I 4/m$ and $Pm3m$ and shows that the former gives a better fit.
The temperature evolution of unit cell volume is presented in Fig.~\ref{xrd} (b).
The variation of cell volume was modelled
using the Gr\"{u}neisen approximation
\cite{wallace1998thermodynamics},
$V (T)$ = $\gamma$ $U (T)$/$K_0$ + $V_0$; $U (T)$ = 9 $Nk_BT(T/\Theta_D)^3$ $\int^{\Theta_D/T}_{0}x^3/(e^x - 1) dx$.
Here, $\gamma$ is the Gr\"{u}neisen parameter,
$K_0$ is the incompressibility, $V_0$ is the volume
at $T$ = 0~K and $\Theta_D$ is the Debye temperature,
and the resulting fit is shown as a solid line
in Fig.~\ref{xrd} (b).
Note that the fit deviates from the data
at $T \sim$ 75~K where the volume
displays a hump-like feature.
This is inferred as magnetoelastic
or magnetovolume effect coinciding with the magnetic transition temperature.
In this respect, further studies aimed at investigating magneto-structural coupling
and polarization measurements would be rewarding.
The value of $\Theta_D$ obtained from the fit is
463(21)~K and $V_0$ = 227.89 \mbox{\AA}$^3$.
The $\Theta_D$ of Sr$_2$FeCoO$_6$ lies in the range of values
estimated for similar double perovskites like Sr$_2$FeTeO$_6$
assuming two different $\Theta_D$'s for light and heavy atoms. \cite{martinjmc_16_66_2006}
An important structural parameter characterizing the 
double perovskites, and having a bearing on the magnetic
properties is the cationic site disorder.
However, since the x-ray atomic structure factors of Fe and Co are
similar, an accurate evaluation of atomic occupancies
is not possible from the present XRD data.
In order to check the degree of $B$ site ordering and
to confirm the crystal structure of Sr$_2$FeCoO$_6$ determined by x rays,
we performed neutron diffraction experiments
(the neutron coherent scattering cross section of Fe and Co are $11.22\times10^{-24}$cm$^2$ and
$0.779\times10^{-24}$cm$^2$ respectively). \cite{nnews}
In Fig~\ref{ND} we present the neutron diffraction pattern of Sr$_2$FeCoO$_6$ collected at 300~K
along with the results of structure refinement using $I4/m$ space group.
Through M\"{o}ssbauer studies \cite{pradheesh},
it has been confirmed that Fe occupies two distinct crystallographic sites;
which supports the choice of $I4/m$ in our structural analysis.
To confirm the nature of $B$ site ordering,
we carried out the refinement of atomic occupancies
using structural models with ordered as well as disordered cationic
arrangement at the $B/B'$ site.
A better fit to the experimental data, in terms of
profile-match and discrepancy factors was
achieved by assuming a disordered model with a
random arrangement for the $B/B'$ cations.
The oxygen content was quantified in the refinement
process and the deviation from stoichiometry was
found to be $\delta \sim$ 0.11.
The structural parameters of Sr$_2$FeCoO$_6$
including atomic positions, lattice parameters and
occupancies and the bond valence sums (BVS) calculated
using VaList program
\cite{ASWills}
are collected in Table~\ref{BVS}.
A rock-salt arrangement of the cations in double perovskites
normally occurs if the charge difference between the
$B$ and $B'$ is greater than 2. \cite{anderson_1993}
In the case of Sr$_2$FeCoO$_6$, the charge difference is close to
unity as clear from Table~\ref{BVS}.
The BVS calculation from the bond length gives
an insight into the oxidation state of the ions 
and reasonably supports random occupancy at
the $B/B'$ site and mixed valence of Fe and Co.
The bond length Fe(2)--O(2) estimated from the
structural analysis is 1.969~\mbox{\AA}
and is comparable to the values found in the literature
for double perovskites, especially Sr$_2$Fe$_{0.75}$Cr$_{0.25}$MoO$_6$
(Fe--O distance is 1.968~\mbox{\AA}). \cite{Blascossc_4_651_2002}
The bond lengths and bond angles obtained from the
analysis of NPD data are given in Table~\ref{BAL}.
According to the Glazer's notation, the space group $I4/m$ has an
antiphase tilting along $c$ axis denoted as $a^0a^0c^-$ which represents
the small shifts of the in-plane oxygen atoms.
\cite{glazer_1972}
The magnitude of tilting can be derived from Fe--O--Co ($\phi$) angle as
$(180-\phi)/2$=$5.65^{\circ}$.
The average bond length of $\langle$Fe1-O$\rangle$ and
$\langle$Fe2-O$\rangle$ are $1.9147~\mbox{\AA}$ and $1.9573~\mbox{\AA}$, where
the latter bond length compare well with the expected value calculated as the
sum of the ionic radii of Fe$^{3+}$ in low spin state (LS)
($0.55~\mbox{\AA}$) and O$^{2-}$ ($1.40~\mbox{\AA}$).
\cite{shannon_1976}
Similarly the bond length values for $\langle$Co1-O$\rangle$
($1.9147~\mbox{\AA}$) and $\langle$Co2-O$\rangle$ ($1.9573~\mbox{\AA}$)
are close to the values of the calculated bond length
if cobalt has a low spin 4+ valence state
($r_{LS}$=$0.518~\mbox{\AA}$)
\cite{taguchi_1979} and an intermediate
spin (IS) 3+ state ($r_{IS}$=$0.56~\mbox{\AA}$)
respectively.
\cite{radaelli_2002}
\\
The magnetization profiles of Sr$_2$FeCoO$_6$
measured at 500~Oe in the zero field-cooled (ZFC) as
well as field-cooled cooling (FCC) and warming (FCW)
cycles are shown in Fig~\ref{MT} (a).
A magnetic transition is evident at $T_{c}\sim$ 75~K.
The low temperature magnetization data
show a marked irreversibility below $T_{c}$ indicating
the presence of a weak component of ferromagnetism as suggested by
Goodenough--Kanamori rules for the $d^4 - d^5$ cation--anion--cation superexchange
\cite{goodenough_100_564_1955,kanamori_10_87_1959}
or spin glass state.
With an increase in applied field, the ZFC/FC curves
tend to merge and the FC arm shows signs of saturation with
broadening of the peak (see inset (1), Fig~\ref{MT} (a)).
This broadening, also evident in inset (2), Fig~\ref{MT} (a) where
the FC magnetization at higher fields are presented,
signifies the presence of majority FM phase since
for an antiferromagnet the increase in the field would have had
little effect on the sharpness of the transition.
\cite{cao_63_184432_2001}
The inverse  magnetic susceptibility in the
temperature range 220 -- 350~K was fitted to
Curie-Weiss law and the result is presented in Fig~\ref{MT} (b) where a
deviation from the Curie-Weiss (CW) law can be clearly seen for $T > T_c$.
Deviation from Curie-Weiss behaviour above $T_c$
attributed to antiferromagnetic spin fluctuations
is observed generally in spin glass systems.
\cite{rao_1983}
Previous reports on SrFe$_{1-x}$Co$_{x}$O$_{3}$ show that
Fe and Co in these solid solutions can possess complex valence states
\cite{munozjssc_179_3365_2006high}
and that, Fe-O-Co superexchange pathways can form which
favour electronic transfer from
Fe$^{3+}$ high spin state (HS, S=$5/2$) to Co$^{4+}$
low spin state (LS, S=$1/2$) via superexchange.
\cite{maignan_3_57_2001}
Complex valence states are observed in the case of Sr$_2$FeCoO$_6$ from
the CW  analysis of magnetic susceptibility and is described in detail below.
The effective paramagnetic moment calculated from
the CW analysis is $\mu_{eff}$ = 3.9~$\mu_{B}$.
Keeping in mind that the system is randomly ordered,
the effective magnetic moment
can be calculated by assuming 50$\%$ of Fe and Co (in both $2a$ and $2b$ sites)
in $4+$ and $3+$ spin states.
The experimentally obtained  effective spin-only moment from the CW fit lies
in between the theoretical spin-only moment values of $\mu_{eff}$ = 3.67~$\mu_{B}$ and
$\mu_{eff}$ = 4.12~$\mu_{B}$ (see, Table~\ref{Spinstate}).
Comparing this with the bond length values obtained we infer that
Co$^{3+}$ is in intermediate spin state while the valence states
of Fe$^{3+}$ and Fe$^{4+}$ along with Co$^{4+}$ are in low spin states.
The positive value of Curie-Weiss temperature, $\Theta_{CW}$ = 99~K, calculated from the fit
suggests the presence of ferromagnetic interactions in the system.
This value is lower than that reported earlier
where the role of oxygen non-stoichiometry on
ordering temperature was discussed.
\cite{munozjssc_179_3365_2006high}
However, oxygen deficient double perovskites exhibit very low
magnetoresistance (MR) which is not the case in Sr$_2$FeCoO$_6$ since we have observed 68$\%$
MR at 12~K at an applied field of 9~T.
\cite{pradheesh}
\\
Fig~\ref{MT} (c) shows the plot of FCC  and FCW arms measured
at 100~Oe where a clear thermal hysteresis below $T_c$ is discernible.
Thermal hysteresis in magnetization cycles are related to
the presence of first-order phase transitions
with mixed magnetic phases as has been reported in the 
case of disordered manganites
\cite{adityajpcm_22_026005_2010,lynn_76_1996}
or rare earth double perovskite oxides.
\cite{dass_67_2003}
The end compositions of Sr$_2$FeCoO$_6$ -- SrFeO$_3$ and SrCoO$_3$  -- possess
helical magnetic structure and FM respectively; hence Sr$_2$FeCoO$_6$ consists of
AFM regions where Fe--O--Fe superexchange dominate
and FM regions where Co--O--Co  which is an ideal scenario for spin glass phase.
Magnetic frustration effects in double perovskite oxides
are also reported to show similar thermal hysteresis below $T_c$
as in the case of Sr$_2M$ReO$_6$ ($M$ = Sc, Co etc).
\cite{katoprb_69_184412_2004}
Drawing parallels, magnetic frustration effects
originating from different exchange paths due to multiple
valences of Co/Fe can also lead to such effects.
We also note that at higher applied fields like 500~Oe
(Fig~\ref{MT} (a)), hysteresis is absent.
These features indicate a weak first-order-like phase transition
due to the presence of mixed magnetic phases.
\\
The isothermal magnetization curve of Sr$_2$FeCoO$_6$ at 2~K is shown in
the main panel, in Fig~\ref{MH} (a) while inset (1)
presents those at 5, 10~K and , 30 - 300~K (inset 2).
Hysteresis is observed below 50~K indicating weak ferromagnetism.
With an increase in temperature, a decrease in  remanance and coercivity 
is observed, see Fig~\ref{MH} (b).
As the temperature increases, the irreversibility reduces and
becomes sigmoidal at 50~K and above; whereas
no saturation is attained even at 50 kOe.
A notable feature in the isothermal magnetization
curves is the step like behaviour in
magnetization at low fields as presented
in the enlarged view in Fig~\ref{MH} (c).
Such a behaviour was observed in Sr$_{2}$YRuO$_{6}$ and has been attributed
to spin flop transition taking place at a critical field.
\cite{cao_63_184432_2001}
However, the aptness of such an explanation in the case of
Sr$_2$FeCoO$_6$ cannot be tested until the magnetic structure is ascertained.
\\
Sharpness of the peak in ac susceptibility curves
and its shift to high temperature
with increasing frequency are typical features exhibited by SG systems.
\cite{tholence_88_1993,mydosh_1993}
In order to probe this, ac susceptibility measurements were performed 
in the temperature range 5 -- 90~K, with a driving field of 3~Oe.
Fig~\ref{acchi} (a) shows the temperature dependence of the real part of ac susceptibility, $\chi'(T)$, 
at different applied frequencies in the range 33 - 1333 Hz.
$\chi'(T)$ attains a maximum  at $T_{f}$, the freezing temperature,
that shifts to higher temperature as the frequency is increased.
This behaviour is  characteristic of SG
\cite{mydosh_1993}
and is clear from Fig~\ref{acchi} (b).
A fit to the Arrhenius relation,
$\omega=\omega_{0}\exp[E_{a}/k_{B}T]$, where $E_a$ is the activation energy,
yielded unphysical values of $\omega_{0}$ = 10$^{82}$ Hz and $E_{a}$ = 14301~eV.
The failure of Arrhenius relation points to SG behaviour, since a 
description of mere energy-barrier blocking and thermal activation will 
not suit the SG transition. 
To confirm the spin glass behaviour, 
analysis of the dynamical scaling at $T_c$ was performed. 
For a spin glass, the critical relaxation time $\tau$ should 
follow a power law divergence of the form,
%
%\begin{equation}\label{tau}
$\tau= \tau_0\,\left(\frac{T_f}{T_{ct} - T_{f}}\right)^{z\nu}$
%\end{equation}
% 
where $T_{ct}	$ is the critical temperature and $z\nu$ and $\tau_0$ 
are the critical exponents and microscopic time scale respectively.
\cite{mydosh_1993}
The best fit to the above equation, as shown in Fig~\ref{acchi} (c)
(inset presents the same graph in log-log scales), yields $z\nu$ = 6.2(2),
$T_{ct}$ = 75.14(8) K and $\tau_{0}\approx10^{-12}$s.
The values of $\tau_{0}$ and $z\nu$
are consistent with that of the conventional 3D spin glasses
\cite{gunnarsson_61_754_1988}
and Fe doped cobaltites.
\cite{luo_75_2007}
In order to probe whether the FM state coexist with the
frozen spin glass state as in {\it reentrant} spin glasses,
we performed ac susceptibility measurements with superimposed dc fields.
The result is presented in Fig~\ref{acchi} (d)
which shows the out of phase susceptibility in
an external dc field varying from 50 - 1000 Oe at a frequency
of 33 Hz with an ac amplitude of 3 Oe.
We do not observe emergence of a second peak
as observed in certain {\it reentrant} systems
\cite{viswanathanprb_80_012410_2009}, instead, 
the intensity of the peak at $T_c$
in $\chi''(T)$ diminishes with increasing field
indicating FM nature of the sample.
The ac susceptibility studies confirm the presence of
canonical SG state in Sr$_2$FeCoO$_6$ in turn suggests
the presence of disorder; since disorder is a key ingredient
for SG behaviour.
\\
To summarize the results, we find that Sr$_2$FeCoO$_6$
crystallizes in tetragonal $I4/m$ where the
$B$ site is randomly occupied by Fe and Co in $2a$ and $2b$ sites,
a finding also supported by the bond valence sums.
The Fe and Co cations exist in mixed valence
states of +3 and +4 in Sr$_2$FeCoO$_6$.
The random crystallographic occupation and mixed valency
sets the stage for mixed FM and AFM interactions between the transition
metal cations and in turn, lead to inhomogeneous
magnetism.
In a perfectly ordered double perovskite, the magnetic
exchange is predominantly governed by the ferromagnetic
Fe$^{3+}$--O$^{2-}$--Co$^{4+}$.
With the occurrence of site disorder, additional
exchange paths are introduced for example, Fe--Fe
or Co--Co which are antiferromagnetic.
The spin glass phase with a $T_c \sim$ 75~K that appears in
Sr$_2$FeCoO$_6$ has it's origin in the multiple exchange
paths that arise due to the mixed interactions.
The magnetic moment at 5~K ($0.39\mu_{B}$)
suggests that the transition metal ions in this compound are
in a mixed valence state unlike a high-spin tetravalent state
as in SrFe$_{0.5}$Co$_{0.5}$O$_3$.
\cite{abbate_2002}
The magnetic ordering of double perovskites is a function of the
strength of the $NN$ and $NNN$ interactions
\cite{battle_1984}
and the competition between these interactions is known to lead to
magnetic frustration effects in Ln$_2$LiRuO$_6$ (Ln = Pr, Nd, Eu, Gd, Tb)
\cite{makowski_2008} and LaBaCoNbO$_6$.
\cite{bos_2004}
Depending on the strength of $NN$ and $NNN$ interactions materials
are classified as Type $\Rmnum{1}$ (strength of AFM interaction is negligible),
Type $\Rmnum{2}$ (dominant $NNN$ interaction)
and Type $\Rmnum{3}$ (strength of AFM interaction significant but less than $NN$ interaction).
Type $\Rmnum{2}$ antiferromagnetic structure leads
to disorder and yields a ferromagnetic $NN$ interaction
along with an antiferromagnetic $NNN$ interaction.
\cite{martinjmc_16_66_2006}
The presence of disorder leading to spin glass behaviour in Sr$_2$FeCoO$_6$
shows that the $NNN$ interaction is not negligible in double perovskite systems.
SG behaviour due to incompatible superexchange
interactions and magnetic frustration was 
shown by the double perovskites,
Sr$_2$FeTiO$_6$
\cite{gibb_2_1992}
and Sr$_2$FeTaO$_6$.
\cite{cussenjmc_7_459_1997}
Since Fe$^{3+}$ and Co$^{4+}$ are isoelectronic,
they interact ferromagnetically through the $NN$
Fe--O--Co superexchange interaction.
However, presence of like-pairs of Fe and Co leads to  Fe--O--Co--O--Fe
$NNN$ antiferromagnetic superexchange interaction also.
\cite{Elad_2003}
The competition between the $NN$ and $NNN$ interactions is the
origin of the magnetic frustration in Sr$_2$FeCoO$_6$ which results in a spin glass state.
One of the structural aspect affecting the strength of
the nearest neighbour interaction is B--O--B$^{\prime}$ angle.
The $NN$ interaction is stronger if the Fe--O--Co bond angle is
near to or equal to $180^{\circ}$ in which case, a ferromagnetic interaction manifests.
A deviation from the linear Fe--O--Co chain in the case of Sr$_2$FeCoO$_6$
was clear from the powder neutron diffraction
analysis where a bond angle equal to $168.98^\circ$ was estimated
which in turn indicates weak $NN$ interaction.
The weak $NN$ interaction also explains the observation of a
lower $T_c$ and the low magnetic moment for Sr$_2$FeCoO$_6$.
\section{Conclusions}
Sr$_{2}$FeCoO$_{6}$  synthesized using sol-gel method 
crystallizes in tetragonal $I 4/m$ structure with disordered
cation occupancy.
Using neutron diffraction studies, we have confirmed
the crystal structure and the subsequent bond valence sums
confirm mixed valence for Fe and Co.
The different valences of Fe and Co and their mixed occupation in the
lattice leads to competition between $NN$ and $NNN$
exchange interactions which is also supported
by the bond angles estimated.
DC magnetization measurements show irreversibility
at $\sim$ 75~K where the spin glass phase sets in.
The weak component of ferromagnetism observed in magnetization
can be due to the competing antiferromagnetic and ferromagnetic interactions and is clear
from the low values of remnance.
The spin glass state in Sr$_2$FeCoO$_6$ is further supported by
the frequency dependence of real component of ac susceptibility.
Dynamical scaling analysis confirms
the spin glass phase with $T_{ct}$ = 75.14(8) K.
We propose that spin glass behaviour is due to the
magnetic frustration resulting from the
competing $NN$ and $NNN$ interactions which originate
from mixed occupation of cations sites.
Our work emphasize that the cationic disorder leads to
a spin glass state.
The structural and magnetic studies also lead to the deduction of the
spin states of Fe and Co in Sr$_2$FeCoO$_6$.
%
%===========================================================================%
\section*{Acknowledgements}
The authors acknowledge the Department of Science and Technology 
(DST), India for the financial support for providing the facilities used in this 
study (Grant No. SR\slash FST\slash PSII-002\slash 2007) and (Grant No. SR\slash NM\slash NAT-02\slash 2005).
PR wishes to thank M. Angst and K. Balamurugan for fruitful discussions.\\
%=================================================================%
$^*$~vsn@physics.iitm.ac.in, ksethu@physics.iitm.ac.in\\
%=================================================================%
%
% FIGURES
%=================================================================%
\clearpage
\newpage

\clearpage
\newpage

\clearpage
\newpage

\clearpage
\newpage

\clearpage
\newpage

\clearpage
\newpage
%
%
%=================================================================%
% Tables
%=================================================================%
%
%
%
\clearpage
\newpage
\begin{table}
\caption{Occupancy and bond valence sums (BVS) and (x,y,z) values 
obtained after Rietveld refinement of the neutron data at 300~K.
The unit cell parameters are $a = 5.4609(2)$\mbox{\AA}, $ b = 5.4609(2)$\mbox{\AA} and $c = 7.7113(7)$\mbox{\AA}.
The quality of the fit is indicated by the discrepancy factors,
$R_{WP}$ = 17.6 $\%$, $R_{P}$ = 14.6 $\%$ and the goodness-of-fit, $\chi^{2}$ = 3.38.}
\footnotesize
\begin{tabular}{llcccll}
%{\textwidth}{@{}l*{15}{@{\extracolsep{0pt plus12pt}}l}}
\hline
Atom         &   Site  &  x           &  y    &   z      &  Occupancy  &   BVS       \\ \hline\hline
Sr           &   $4d$  & $0$          & $0.5$ &  $0.25$  &  $0.25$     & $2.336(9)$   \\
Fe1/Co1      &   $2a$  & $0$          & $0$   &  $0$     &  $0.0625$   & $4.055(69)/3.650(62)$\\
Fe2/Co2      &   $2b$  & $0$          & $0$   &  $0.5$   &  $0.0625$   & $3.51(41)/2.967(54)$\\
O1           &   $4e$  & $0$          & $0$   &  $0.24739(3)$ & $0.25994(10)$ & $1.970(38)$\\
O2           &   $8h$  & $0.26798(13)$&$0.21981(18)$ & $0$ & $0.47621(6)$ & $1.985(15)$\\ \hline\hline
\label{BVS}
\end{tabular}
\end{table}
\clearpage
\newpage
\begin{table}
\caption{Selected bond angles and bond lengths of Sr$_2$FeCoO$_6$ obtained from the structural
analysis.}
\footnotesize
\begin{tabular}{ll}
%{\textwidth}{@{}l*{15}{@{\extracolsep{0pt plus12pt}}l}}
\hline
Bond length (\mbox{\AA}) Bond angle (deg) &  \\ \hline\hline
%\multicolumn{2}{|c|}{Bond Length and Bond Angle from the refinement} \\
%\hline
%\hline
Fe1/Co1-O1        & $1.9227(15)$$\times2$ \\
Fe1/Co1-O2        &$1.9107(15)$$\times4$   \\
Fe2/Co2-O1        &$1.9329(15)$$\times2$      \\
Fe2/Co2-O2        &$1.9695(15)$$\times4$       \\
Fe1-O2-Co2    &$168.6988(5)$                  \\ \hline\hline
\label{BAL}
\end{tabular}
\end{table}
\clearpage
\newpage
\begin{table}
\caption{Calculated spin only moments ($\mu^2_{SO}$=$\mu^2_{2a}$+$\mu^2_{2b}$) for Sr$_2$FeCoO$_6$ assuming the possible
spin state values. The calculated spin only moment ($\mu_{eff.}$) lies in between the last two configurations.
Note that Co$^{3+}$ HS $S=2$, LS $S=0$, IS $S=1$ ; 
Co$^{4+}$ HS $S=5/2$, LS $S=1/2$ ;
Fe$^{3+}$ HS $S=5/2$, LS $S=1/2$ ; Fe$^{4+}$ HS $S=2$, LS $S=1$. }
\footnotesize
\begin{tabular}{ccc}
%{\textwidth}{@{}l*{15}{@{\extracolsep{0pt plus12pt}}l}}
\hline
2a                                     &  2b                                      & $\mu_{SO} (\mu_{B})$                 \\ \hline\hline
Co$^{4^+}$$_{HS}$($S=5/2$); Fe$^{4^+}$$_{HS}$($S=2$) &\hspace{-4mm}Co$^{4^+}$$_{HS}$($S=5/2$); Fe$^{4^+}$$_{HS}$($S=2$)   & $7.68$            \\
Co$^{3^+}$$_{HS}$($S=2$)\hspace{2mm} ; Fe$^{4^+}$$_{HS}$($S=2$) & Co$^{4^+}$$_{HS}$($S=5/2$); Fe$^{4^+}$$_{HS}$($S=5/2$) & $7.68$             \\
Co$^{4^+}$$_{HS}$($S=5/2$); Fe$^{4^+}$$_{LS}$($S=1$) &\hspace{-4mm}Co$^{4^+}$$_{HS}$($S=5/2$); Fe$^{4^+}$$_{LS}$($S=1$)   & $6.56$            \\
Co$^{4^+}$$_{LS}$($S=1/2$); Fe$^{4^+}$$_{HS}$($S=2$) &\hspace{-4mm}Co$^{4^+}$$_{LS}$($S=1/2$); Fe$^{4^+}$$_{HS}$($S=2$)   & $5.19$            \\
Co$^{4^+}$$_{LS}$($S=1/2$); Fe$^{4^+}$$_{LS}$($S=1$) &\hspace{-4mm}Co$^{4^+}$$_{LS}$($S=1/2$); Fe$^{4^+}$$_{LS}$($S=1$)   & $3.32$            \\
Co$^{4^+}$$_{LS}$($S=1/2$); Fe$^{4^+}$$_{LS}$($S=1$) &\hspace{-4mm}Co$^{3^+}$$_{IS}$($S=1$)\hspace{2mm} ; Fe$^{4^+}$$_{LS}$($S=1$)   & $3.67$            \\
Co$^{3^+}$$_{IS}$($S=1$)\hspace{2mm} ; Fe$^{4^+}$$_{LS}$($S=1$)  &\hspace{-4mm}Co$^{4^+}$$_{IS}$($S=3/2$); Fe$^{3^+}$$_{LS}$($S=1/2$) & $4.12$ \\\hline\hline     
\label{Spinstate}
\end{tabular}
\end{table}
\clearpage
\newpage
% ================================================================%
%\section*{References}
%=================================================================%

%=================================================================%
%\newpage
%\bibliography{spinglassSFCO}
%\bibliographystyle{apsrev}
%\nocite{*}
%\bibliography{spinglassSFCO}
%\bibliographystyle{iopart-num}% Produces the bibliography via BibTeX.

\end{document}